\documentclass[a4paper]{mem}
\usepackage{natbib}
\usepackage{graphicx,psfig}
\usepackage[a4paper]{hyperref}
\idline{73}{23}
\begin{document}
   \title{Stellar Abundances in Young and Intermediate Age GCs
}

   \author{Uta Fritze -- v. Alvensleben}

   \institute{Universit\"atssternwarte G\"ottingen, 
Geismarlandstr. 11, 37083 G\"ottingen, Germany, \email{ufritze@uni-sw.gwdg.de}
             }

   \abstract{Globular Cluster ({\bf GC}) formation seems to be a widespread mode of star formation in extreme starbursts triggered by strong interactions and mergers of massive gas-rich galaxies. We use our detailed chemically consistent evolutionary synthesis models for spiral galaxies to predict stellar abundances and abundance ratios of those second generation GCs as a function of their age or formation redshift. Comparison with observed spectra of young star clusters formed recently in an ongoing intercation (NGC 4038/39) and a merger remnant (NGC 7252) are encouraging. Abundances and abundance ratios (and their respective spreads) among young and intermediate cluster populations and among the red peak GCs of elliptical/S0 galaxies with bimodal GC color distributions are predicted to bear a large amount of information about those clusters' formation processes and environment. Not only the bright young clusters but also representative populations of "old" GCs in E/S0 galaxies are readily accessible to MOS on 10m class telescopes.
   
     \keywords{GC abundances --
                GC formation in starbursts --
                 GC formation in galaxy mergers}
               
   }
   \authorrunning{U. Fritze -- v. Alvensleben}
   \titlerunning{Abundances in Young and Intermediate Age GCs}
   \maketitle
%

\section{Introduction}
Evolutionary Synthesis ({\bf ES}) models describe the time evolution of galaxy properties on the basis of average Star Formation Histories ({\bf SFH}) appropriate for the respective galaxy types. We have developed a combined chemical and spectral ES code that allows for a {\bf chemically consistent} description of galaxies. We use 5 sets of input physics (stellar lifetimes and yields, evolutionary tracks or isochrones, and model atmosphere spectra) for metallicities [Fe/H]$=-1.7,~-0.7,~-0.4,~0,~+0.4$ and account for the increasing metallicity of successive stellar generations by using input physics appropriate for their respective initial metallicities as given by the gas phase metallicity at their birth. For a standard Salpeter initial mass function the SFHs for Sa, Sb, Sc, Sd models are strongly constrained by average observed colors U . . . K, template spectra, emission line strengths, and characteristic HII region abundances. 

\section{Star Cluster Formation in Galaxy Mergers}
We use the evolution of spiral ISM abundances as presented in \citet{L99} to predict the metallicities of star clusters forming in spiral-spiral mergers at various evolutionary times or redshifts (cf. Fig. 1). 

   \begin{figure}
   \centering
   \includegraphics[width=\columnwidth]{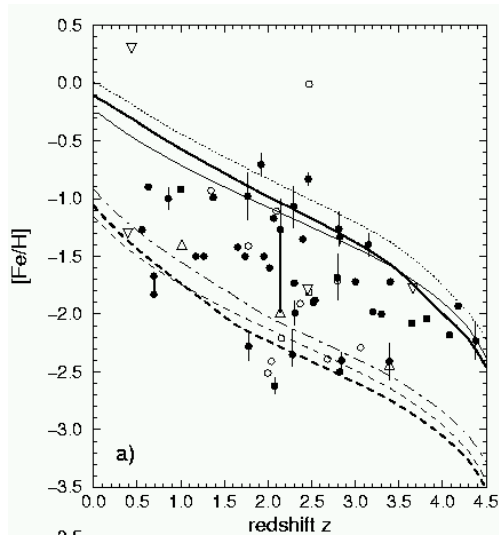}
      \caption{Redshift evolution of [Fe/H] for Sa (thick upper line) and Sd (thick lower line) models for ${\rm (H_o=50,~\Omega_o=1)}$ from \citet{L99} (with symbols denoting Damped Lyman Alpha absorber abundances).}
   \end{figure}

\section{Abundances of Young Globular Clusters}
Young globular clusters forming in spiral-spiral mergers will to first order 
\begin{itemize}
\item form out of ISM pre-enriched in spirals
\item with abundance ratios [$\alpha$/Fe] $\sim$ solar due the to long timescales for SF in spirals (${\rm \tau_{SF} \gg 10^9}$ yr)
\end{itemize}
GCs forming relatively late during the extended bursts may eventually 
\begin{itemize}
\item show additional enrichment during the bursts (typically ${\rm \tau_b \sim 10^8}$ yr)
\item and abundance ratios [${\rm \alpha/Fe] > 0}$
\end{itemize}
in fair agreement with young GC spectra in NGC 7252 (\citealt{SS93}, \citealt{SS98}, \citealt{FB95}). 

\section{Abundances of Intermediate Age Globular Clusters} 
GCs from the red peak of the often bimodal GC color distributions in E/S0s are good candidates for intermediate age GCs.
\begin{itemize}
\item red peak GCs in E/S0 galaxies show ${\rm \langle V-I \rangle \sim 1.2}$
\item their ages and metallicities bear information about their host galaxy and its violent formation processes
\end{itemize}
Calculating the color evolution of star clusters of various metallicities (\citealt{S02}, \citealt{AF03}), we find that GCs with ${\rm \langle V-I \rangle \sim 1.2}$ can have a wide range of age-metallicity combinations with very different respective ${\rm V-K}$ colors: from ${\rm [Fe/H]=-0.4 ~and~ 13}$ Gyr with ${\rm V-K=3.1}$ to ${\rm [Fe/H]=+0.4 ~and ~3}$ Gyr with ${\rm V-K=3.6}$. 
We predict that ${\rm V-K}$ data will allow to considerably narrow down the range of age-metallicity combinations. MOS on 8-10 m class telescopes, feasible up to Virgo cluster distances, will soon yield precise abundances and abundance ratios, and hence ages, for these GCs. 
\begin{acknowledgements}
      I gratefully acknowledge partial travel grants from the DFG (Fr 916/12-1) and from the IAU.
\end{acknowledgements}

\bibliographystyle{aa}

\end{document}